\documentclass[a4paper]{jpconf}
\usepackage{graphicx}
\def\msun{$M_{\odot}$}

\begin{document}
\title{Modeling He-rich subdwarfs through the hot-flasher scenario}

\author{M. M. Miller Bertolami$^{1,2}$,
        L. G. Althaus$^{1,2}$,
	K. Unglaub$^3$,
	A. Weiss$^{4}$}
\address{$^1$Facultad de Ciencias  Astron\'omicas y Geof\'{\i}sicas,
           Universidad Nacional de La Plata,\\ Paseo del Bosque s/n,
           (1900) La Plata, Argentina.}  
\address{$^2$Instituto de
           Astrof\'{\i}sica La Plata, UNLP-CONICET}  
\address{$^3$Dr. Remeis-Stenwarte Bamberg, Sterwartstr. 7, D96049
           Bamberg, Germany}
\address{$^4$Max-Planck-Institut f\"ur
           Astrophysik, Karl-Schwarzschild-Str. 1, 85748, Garching,
           Germany.}
\ead{mmiller@fcaglp.unlp.edu.ar} 

\begin{abstract}
We present 1D numerical simulations aimed at studying the hot-flasher
scenario for the formation of He-rich subdwarf stars. Sequences were
calculated for a wide range of metallicities and with the He core
flash at different points of the post-RGB evolution (i.e. different
remnant masses).  We followed the complete evolution from the ZAMS,
through the hot-flasher event, and to the subdwarf stage for all kinds
of hot-flashers. This allows us to present a homogeneous set of
abundances for different metallicities and all flavors of
hot-flashers. We extend the scope of our work by analyzing the effects
in the predicted surface abundances of some standard assumptions in
convective mixing and the effects of element diffusion.

We find that the hot-flasher scenario is a viable explanation for the
formation of He-sdO stars. Our results also show that element
diffusion may produce the transformation of (post hot-flasher) He-rich
atmospheres into He-deficient ones. If this is so, then the
hot-flasher scenario is able to reproduce both the observed properties and
distribution of He-sdO stars.

\end{abstract}

\section{Introduction}

Hot subdwarf stars can roughly be grouped into the cooler sdB stars,
whose spectra display typically no or only weak helium lines, and the
hotter sdO stars, which have higher helium abundances on average and
can even be dominated by helium. The diversity of spectra observed
among sdO stars and the helium-enhanced surface abundances observed in
many of these stars (Lemke et al. 1997) pose a challenge to stellar
evolution theory. As a consequence, some non-canonical evolutionary
scenarios were proposed to explain the formation of He-rich subdwarf
stars (i.e. He-sdB, He-sdO stars). Among them, the merger of two white
dwarfs (Saio \& Jeffery 2000) and a helium core flash (HeCF) after
departure from the red giant branch (i.e. ``hot-flasher scenario'';
D'Cruz et al. 1996, Sweigart 1997) offer the most promising
explanations of their formation (Str\"oer et al. 2007). The
hot-flasher scenario was also proposed to explain the existence and
characteristics of blue hook stars in some globular clusters (see Moehler
et al. 2007 and references therein). Following Lanz et al. (2004)
hot-flashers can be classified into \emph{early} hot-flashers, which
experience the HeCF during the evolution at constant luminosity (after
departure from the RGB) and become hot subdwarfs with standard H/He
envelopes; and \emph{late} hot-flashers, which end as He-enriched hot
subdwarfs either by burning or dilution of the remaining H-rich
envelope. The latter case is the object of the present article.

\section{Description of the present work}

The core feature of this work is to present a homogeneous set of
simulations of the {\emph late} hot-flasher scenario (i.e. those
events that lead to He-enhanced stellar surfaces) for which the
chemical abundances are consistently calculated by simultaneously
solving the mixing and burning equations within a diffusive convection
picture. We then extend the scope of this work by analyzing how these
results can be altered by some non-standard assumptions, such as
deviations from the standard mixing length theory (MLT) or element
diffusion. Particular emphasis is placed on on the predicted surface
properties of the models.

The sequences presented in this work were calculated with {\tt
LPCODE}, a numerical code for solving the equations of stellar
structure which is extensively described in Althaus et al. (2005). 
 
 In our simulations we have followed the evolution from the ZAMS,
through the hot-flasher event and then to the He-core burning
stage. Simulations of the hot-flasher scenario were performed for 4
different metallicities and with the HeCF taking place at different
points of the post-RGB evolution (which corresponds to different
remnant masses, see Table 1).  Initial masses and abundances are shown
on Table 1 together with the range of post-RGB remnant masses for
which He-enriched subdwarfs are obtained.

\begin{table}
\begin{center}
{\small
\begin{tabular}{ccc}
Mass at ZAMS & Initial Abundances & Remnant mass range \\ 
(\msun) & (X$_0$/Y$_0$/Z$_0$) &for \emph{late} hot-flashers (\msun) \\ \hline 
0.88 & 0.769/0.230/0.001 & 0.4916 --- 0.4810 \\ 
0.98 & 0.736/0.254/0.010 & 0.4780 --- 0.4662 \\ 
1.03 & 0.702/0.278/0.020 & 0.4740 --- 0.4611 \\
1.04 & 0.668/0.302/0.030 & 0.4658 --- 0.4522 \\
\end{tabular}
\label{tab:inicial}
\caption{Initial (ZAMS) values of our standard sequences and range of
remnant masses in which \emph{late} hot-flashers are obtained.}
}
\end{center}
\end{table}

For a more extensive discussion on the numerical and physical aspects
of the models, as well as on the results of the present study, we
refer the reader to Miller Bertolami et al. (2008).
\section{Results and discussion}
\begin{table*}[ht!]
\begin{center}
{\small
\begin{tabular}{c|c||c|c|c|c|c|c}
Final Mass [\msun]& Z$_0$        & H   & He &  $^{12}$C  &  $^{13}$C
&  N   &  O   \\ \hline\hline
0.49145 (SM$^*$)& 0.001   & 0.2137                & 0.7475 & 0.0365 & $1.57
\times 10^{-6}$ & $1.22 \times 10^{-4}$ & $1.52 \times 10^{-3}$ \\
0.49104 (DM)    & 0.001   & $2.42 \times 10^{-4}$ & 0.9450 & 0.0267 & $7.47
\times 10^{-3}$ & 0.0106 & $4.77 \times 10^{-5}$  \\
0.48545 (DM)    & 0.001   & $4.60 \times 10^{-6}$ & 0.9524 & 0.0264 & $7.73
\times 10^{-3}$ & 0.0134 & $4.53  \times 10^{-5}$ \\
0.48150 (DM)    & 0.001   & $2.41\times 10^{-6}$ & 0.9666 & 0.0107 & $5.79
\times 10^{-3}$  & 0.0185 & $ 3.79\times 10^{-5}$ \\\hline
0.47770 (SM)    & 0.01 & 0.0780 & 0.8682 & 0.0426 & $6.26 \times 10^{-6}$  &
$1.30\times 10^{-3}$ & $6.84\times 10^{-4}$ \\
0.47744 (SM$^*$) & 0.01 & 0.0261 & 0.8996 & 0.0642 & $2.36 \times 10^{-6}$  &
$6.59\times 10^{-4}$ & $7.43\times 10^{-4}$ \\
0.47725 (DM)    & 0.01 & $2.27\times 10^{-5} $ & 0.9464 & 0.0382 & $6.12\times
 10^{-3}$ & $4.35 \times 10^{-3}$ & $2.94\times 10^{-4}$ \\
0.46921 (DM)    & 0.01 & $1.02\times 10^{-5} $ & 0.9560 & 0.0211 & $6.13\times
10^{-3}$ & 0.0128 & $2.92\times 10^{-4}$ \\
0.46644 (DM)    & 0.01 & $4.61 \times 10^{-6}$ & 0.9592 & 0.0193 & $5.52\times
10^{-3}$ & 0.0122 & $3.16 \times 10^{-4}$ \\\hline
0.47378 (SM)     & 0.02 & 0.3822 & 0.5978 & $1.23 \times 10^{-3}$ & $8.96
\times 10^{-5}$ &  $7.57 \times 10^{-3}$ & $4.69 \times 10^{-3}$ \\
0.47250 (SM$^*$) & 0.02 & $9.56 \times 10^{-3}$  & 0.9273 & 0.0425 & $1.93
\times 10^{-5}$ & $2.94 \times 10^{-3}$ & $8.64 \times 10^{-4}$ \\
0.47112 (DM)     & 0.02 & $9.05 \times 10^{-6}$  & 0.9389 & 0.0420 & $ 4.27
  \times 10^{-3}$ & $4.88 \times 10^{-3}$ & $6.96 \times 10^{-4}$ \\
0.46410 (DM)     & 0.02 & $1.05\times 10^{-5}$   & 0.9460 & 0.0273 &
$6.41\times 10^{-3}$ & 0.0120 & $7.680 \times 10^{-4}$ \\
0.46150 (DM)     & 0.02 & $4.66\times 10^{-6}$   & 0.9480 & 0.0263 &
$5.67\times 10^{-3}$ & 0.0116 & $1.01  \times 10^{-3}$  \\\hline
0.46521 (SM)    & 0.03    & 0.2227                & 0.7271 & 0.0197 & $7.79
\times 10^{-5}$ & $8.09\times 10^{-3}$ & $5.18 \times 10^{-3}$ \\
0.46470 (SM)    & 0.03    & $4.71 \times 10^{-3}$ & 0.9239 & 0.0380 &
$4.12\times 10^{-4}$ & $6.10\times 10^{-3}$ & $1.58  \times 10^{-3}$ \\
0.46367 (SM$^*$) & 0.03   &  $1.63 \times 10^{-3}$ & 0.9249 & 0.0390 &
$3.25\times 10^{-3}$ &  $6.48\times 10^{-3}$ & $1.62  \times 10^{-3}$ \\
0.46282 (DM)    & 0.03   &  $7.48 \times 10^{-6}$ & 0.9333 & 0.0420 &
$1.74\times 10^{-3}$ &  $7.04\times 10^{-3}$ & $1.55\times 10^{-3}$  \\
0.45660 (DM)    & 0.03   &  $9.22 \times 10^{-6}$ & 0.9382 & 0.0341 &
$4.09\times 10^{-3}$ & 0.0103 &$1.67\times 10^{-3}$  \\
0.45234 (DM)    & 0.03   &  $1.94 \times 10^{-5}$ & 0.9411 & 0.0273 &
 $4.76\times 10^{-3}$ & 0.0134 &$2.14\times 10^{-3}$
\end{tabular}
\label{tabla_2}
\caption{Surface abundances (by mass fractions) for sequences calculated in
this work.}
}
\end{center}
\end{table*}

{\bf Standard sequences:} Convection in our standard sequences is
described by the standard MLT with convective boundaries given by a
bare Schwarzschild criterium.  H-deficiency in our standard sequences
is attained by dilution and/or burning of the remaining H-rich
envelope, depending on the exact moment of the post-RGB evolution at
which the HeCF takes place.  If the star is already at the WD cooling
track when the HeCF develops ({\it deep mixing} events; DM) then the
convective zone generated by the HeCF is able to penetrate into the
H-rich envelope and H is violently burned (a H-flash), before the
remaining H-envelope is diluted by a very shallow convective zone.  On
the other hand, if the H-burning shell is still active when the HeCF
develops then, the convective zone generated during the HeCF does not
reach the H-rich envelope and no H-burning happens. In these cases
({\it shallow mixing} events; SM) a mild H-deficiency can be
attained when one of the convective zones generated by the HeCF moves
towards the H-He transition and merges with a inwardly growing
convective envelope. In our simulations there are some transition
cases (labelled SM$^*$ in Table 2) in which the HeCF driven convective
zone partially penetrates into the H-rich envelope, and some H-burning
takes place, but no H-flash finally developes. Then H-deficiency is
attained by a combination of burning and dilution of the H-rich
envelope.

In Table 2 we present the surface abundances for selected sequences
calculated in this work. As it is shown in Table 2, DM episodes
predict {\it immediately after the flash} an extremely H deficient
surface with H-abundances preferentially between $10^{-5}$ and
$10^{-6}$ (by mass fraction), while SM/SM$^*$ predict a more wide
range of H-abundances between almost solar and  $10^{-3}$.

Although evolutionary tracks pass close to the location of He-sdO
stars (see Figure 1), their distribution in the $\log g-\log T_{\rm
eff}$ diagram is not correctly reproduced. In fact, according to
(standard) theoretical predictions, He-sdO should cluster around the
He-core burning stage where evolutionary timescales are longer. However, this
inconsistency can be solved when taking into account the effects of
element diffusion in the outer layers (see next paragraphs).
%\section{Discussion of non standard effects}

{\bf Effects of departures from the standard MLT}: we have performed
additional simulations including the effects of extramixing at
convective boundaries (as in Herwig et al. 1997) and/or the effects of
chemical gradients (as in Grossman \& Taam 1996). Our results show
that the role of chemical gradients as an extra barrier to the
penetration of the He-core flash driven convective zone into the
H-rich envelope, is significant and has to be studied in detail
(probably by means of hydrodynamical simulations of the event). In
fact, in our simulations, the stabilizing effect of chemical gradients
can even prevent the penetration when no extramixing is considered at
convective boundaries. On the other hand, extramixing processes alone
does not produce significant departures from standard results.

{\bf Effects of element diffusion on the outer layers:} Recently,
Unglaub (2008) showed that chemical homogeneous winds in these stars
are not possible at $\log g > 5.5$ and, consequently, H and He can
hardly be expelled from the stars. In view of these results we have
performed simulations of element diffusion in the absence of mass loss
by means of the method described in Unglaub \& Bues (1998). Figure 1
shows the evolution of the surface abundaces predicted for two of
our sequences due to the effects of element diffusion. As it is shown
in Figure 1 in the absence of mass loss the SM$^*$ sequence becomes
H-dominated almost instantly, compared to the evolutionary timescale,
and its surface is expected to become H-dominated as soon as
homogeneous winds halt. On the other hand for the DM sequence the
timescale in which the star becomes H-dominated is similar to the
evolutionary timescale from the primary HeCF to the settlement on the
He-core burning stage ($\sim 1$ Myr). In this case, the surface is
expected to become H-dominated as soon as the He-core burning stage
begins and evolutionary timescales become $\sim 40$ times longer. Note
that, with the inclusion of element diffusion, the surface properties
and distribution of He-sdO stars predicted by the hot-flasher scenario
correctly reproduces the observed ones (see Figure 1).

\begin{figure}[t]
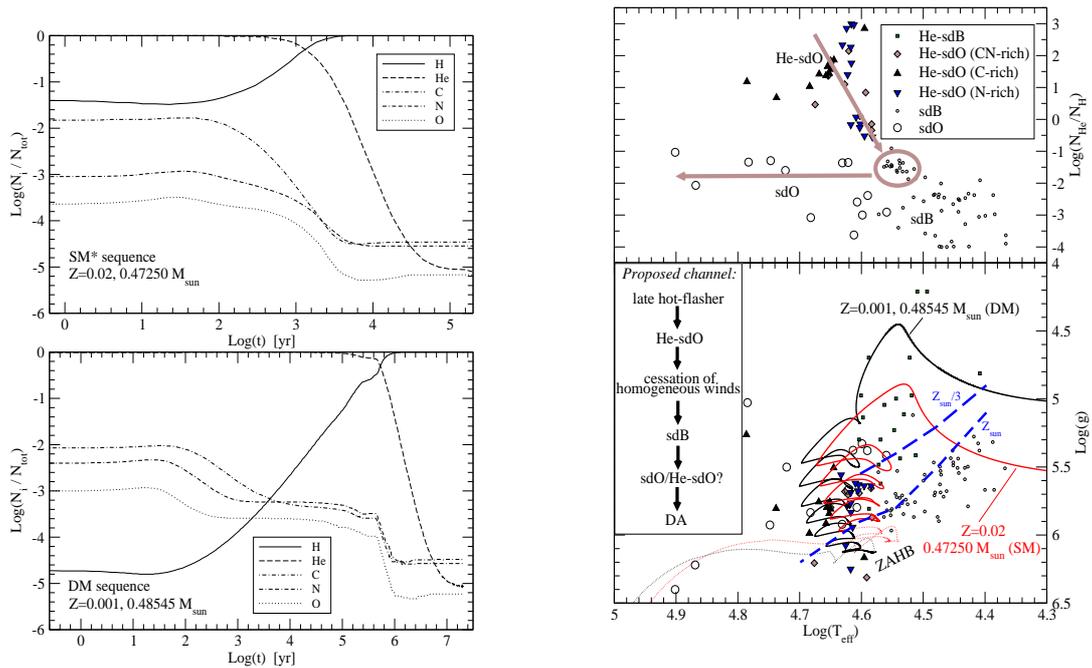

\vskip 8.5cm
\includegraphics{chem.eps}
\includegraphics{g-teff.eps}
\caption{\small {\it Left:} Evolution of the surface ($\tau_{\rm
Rosseland}=2/3$) abundances predicted in two of our hot-flashers due
to the effects of element diffusion in the absence of mass loss and
assuming $T_{\rm eff}=40\,000$ K and $\log g =6$. {\it Bottom Right:}
Evolution of the same two hot-flasher sequences in the $\log g-\log
T_{\rm eff}$ diagram. Thick dashed lines mark the wind limit, for two
different metallicities, below which homogeneous winds are not possible
according to Unglaub (2008). Dotted lines denoted the portion of the
evolutionary track where the star should have already became
H-deficient according to element diffusion simulations. {\it Top
Right:} Sketch of the possible evolution in the $\log N_{\rm
He}/N_{\rm H}-\log T_{\rm eff}$ diagram after a deep mixing event. }
\end{figure}

\ack {\small Part of this work was supported by PIP 6521 grant from
CONICET.  Also the European Association for Research in Astronomy is
acknowledged for an EARA-EST fellowship under which this work was
started.

\section*{References}

\begin{thereferences}

%\item Ahmad, A., Behara, N., Jeffery, C., Sahin, T. \& Woolf, V. 2007, A\&A, 465, 541
\item Althaus, L. G., Serenelli, A. M.,  Panei, J. A., et al. 2005,
{\sl A\&A}, {\bf 435}, 631
%\item Brown, T., Sweigart, A., Lanz, T., Landsman, W.,
%  Hubeny, I. 2001, ApJ, 562, 368
%\item Castellani, M., Castellani, V., \& Prada Moroni, P. 2006, A\&A, 457, 569
\item Cassisi, S., Schlattl, H., Salaris, M., Weiss,  A. 2003, {\sl ApJL}, {\bf 582}, L43
\item D'Cruz, N., Dorman, B., Rood, R., O'Connell, R. 1996,
  {\sl ApJ}, {\bf 466}, 359
\item de Boer, K., Drilling, J., Jeffery, C. \& Sion,
  E. 1997, Third Conference on Faint Blue Stars, L. Davis Press, 515
%\item Grossman, S, Narayan, R. \& Arnett, D. 1993, ApJ, 407, 284
\item Grossman, S \& Taam, R. 1996, {\sl MNRAS}, {\bf 283}, 1165
\item Herwig, F., Bl\"ocker, T., Sch\"onberner, D. \& El Eid,
  M. 1997, {\sl A\&A}, {\bf 324}, L81
%\item Herwig, F., Freytag, B, Hueckstaedt, R. \& Timmes, F. 2006,
%  ApJ, 642, 1057
%\item Herwig, F., Freytag, B, Fuchs, T., Hansen, J., Hueckstaedt,
%  R., Porter, D., Timmes, F. \& Woodward, P. 2007,   ASPC, 378, 43
%\item Hirsch, H., Heber, H. and O'Toole, S. 2008, ASP Conference
%  Series, 392, 175
%\item Iben, I. \& Mac Donald, J. 1995, Lecture Notes in Physics, 443, Springer-Verlag, Berlin, 48
%\item Jeffery, C. S. 2008,  ASP Conference
%  Series, 391, 3
\item Lanz, T., Brown, T., Sweigart, A., Hubeny, I.,
  Landsman, W. 2004, {\sl ApJ}, {\bf 602}, 342
%\item Lemke, M., Heber, U., Napiwotzki, R., Dreizler, S. \&
%  Engels, D. 1997, Third Conference on Faint Blue Stars, L. Davis Press, 375
\item Lemke, M., Heber, U., Napiwotzki, R., et al. 1997, Third
Conference on Faint Blue Stars, L. Davis Press, 375
%\item Lisker, T., Heber, U., Napiwotzki, R., Christlieb, N, Han,
%Z., Homeier, D. \& Reimers, D. 2005, A\&A, 430, 223
%\item Michaud, G., Bergeron, P., Heber, U. \& Wesemael, F. 1989, ApJ 338, 417
\item Miller Bertolami, M. M., Althaus, L. G., Unglaub, K.,
\& Weiss, A. 2008, {\sl A\&A}, DOI: 10.1051/0004-6361:200810373 
%\item  Moehler, S., Sweigart, A., Landsman, W. and  Dreizler,
%  S. 2002, A\&A, 395, 37
%\item   Moehler, S., Sweigart, A., Landsman, W., Hammer, N., and
%  Dreizler, S. 2004, A\&A, 415, 313
%\item Moehler, S., Dreizler, S., Lanz, T., Bono, G, Sweigart, A.,
%  Calamida, A., Monelli, M., and Nonino, M. 2007, A\&A, 475, L5
\item Moehler, S., Dreizler, S., Lanz, T., et al. 2007, {\sl A\&A},
{\bf 475}, L5
\item Saio, H. \& Jeffery, S. 2000, {\sl MNRAS}, {\bf 313}, 671
%\item Str\"oer, A., Heber, U., Lisker, T., Napiwotzki, R.,
%  Dreizler, S., Christlieb, N., Reimers, D. 2007, A\&A, 462, 269
\item Str\"oer, A., Heber, U., Lisker, T., et al. 2007, {\sl A\&A},
{\bf 462}, 269
\item Sweigart, A. 1997, Third Conference on Faint Blue Stars,
L. Davis Press, 3
%\item Unglaub, K. 2005, ASP Conference Series, 334, 297
\item Unglaub, K. 2008, {\sl A\&A}, {\bf 486}, 923
\item Unglaub, K. \& Bues, I. 1998, {\sl A\&A}, {\bf 338}, 75
%\item Unglaub, K. \& Bues, I. 2000, A\&A, 359, 1042
%\item Unglaub, K. \& Bues, I. 2001, A\&A, 374, 570
%\item Vink, J. \& Cassisi, S. 2002, A\&A, 392, 553
%\item Woodward, P., Herwig, F., Porter, D., Fuchs, T., Nowatzki,
%  A., Pignatari, M. 2008, AIPC, 990, 300
\end{thereferences}

\end{document}